\documentclass{emulateapj}

\usepackage{graphicx}
\usepackage{amssymb}

\usepackage{color}
\usepackage{ulem}
\definecolor{Green}{rgb}{0.25,0.85,0.}

\shorttitle{Gravitational waves in black-hole forming core collapse}
\shortauthors{Cerda-Duran et al.}

\begin{document}

\title{Gravitational wave signatures in black-hole-forming core collapse}

\author{Pablo Cerd\'a-Dur\'an, Nicolas DeBrye, Miguel A Aloy,
Jos\'e A Font and Martin Obergaulinger }
\affil{Departamento de Astronomia y Astrof\'isica, Universidad de Valencia, c/ Dr. Moliner, 50, 46100 - Burjassot, Spain}
\email{pablo.cerda@uv.es}

\begin{abstract}
  We present numerical simulations in general relativity of collapsing
  stellar cores. Our initial model consists of a low metallicity
  rapidly-rotating progenitor which is evolved in axisymmetry with the
  latest version of our general relativistic code {\tt CoCoNuT}, which
  allows for black hole formation and includes the effects of a
  microphysical equation of state (LS220) and a neutrino leakage scheme
  to account for radiative losses.  The motivation of our study is to
  analyze in detail the emission of gravitational waves in the collapsar
  scenario of long gamma-ray bursts. Our simulations show that the phase
  during which the proto-neutron star (PNS) survives before ultimately
  collapsing to a black hole is particularly optimal for gravitational
  wave emission. The high-amplitude waves last for several seconds and
  show a remarkable quasi-periodicity associated with the violent PNS
  dynamics, namely during the episodes of convection and the
  subsequent nonlinear development of the standing-accretion shock
  instability (SASI). By analyzing the spectrogram of our simulations we are able
  to identify the frequencies associated with the presence of g-modes and with the 
  SASI motions at the PNS surface. We note that the gravitational waves emitted reach large
  enough amplitudes to be detected with third-generation detectors as
  the Einstein Telescope within a Virgo cluster volume at rates
  $\lesssim 0.1\,$y$^{-1}$.
\end{abstract}



\section{Introduction}

The collapse of the core of a massive and rotating star is the most
likely mechanism to generate central engines suitable to explain the huge
energy output inferred from long lasting gamma-ray bursts (GRBs)
\citep{Woosley93}. One of the theoretical requisites to produce a viable
engine for GRBs is that a central black hole (BH) forms with a mass $\ge
3\,M_\odot$ surrounded by a thick accretion disk. During the formation
of the BH-accretion disk system, a low-density funnel shall also form
along the rotational axis of the system, to provide a suitable
least-resistance region for any ultrarelativistic outflow to escape
without accumulating an excessive amount of baryonic mass.  The path
followed by the collapsing core until it forms a GRB central engine was
early devised by \citet{Macfadyen99} and \citet{MWH01}. More recently, \citep{OO11,Debrye13}
have argued that BH formation in a massive stellar core is almost always
preceded by the formation of a proto-neutron star (PNS) 
which may last for $\sim 0.8$~s. 

The formation of BHs in massive stars is typically associated with
electromagnetic counterparts which are dimmer than typical core collapse
supernovae (SNe) \citep{HBK11,WH12,Kochanek13}, and with failed SNe or
unnovae \citep{LW13,Piro13}. \citet{ZWH08} estimate the fraction of
supernova-like compact remnants that can be produced by fall back
mechanisms, finding that the least energetic explosions experience more
fall back and, hence, produce more massive remnants. Indeed,
\citet{ZWH08} estimate that, leaving aside any influence of rotation and
binary membership of the progenitor star, the fraction of remnants that
will be BHs can be up to $\sim 90\%$. In a
more sophisticated numerical setting, \citet{Uetal12} conclude that $\sim
23\%$ of the potential massive progenitors with solar-metallicity end up
as BHs. \citet{Fetal12} point out that it is difficult to make a strong
SN explosion if there is a long delay between shock formation and the
convective phase of typical convection-enhanced, neutrino-driven
engines. If the explosion occurs less than 250\,ms after bounce, SN
energies in excess of $10^{51}\,$erg are expected.

In this work we consider the process of BH formation by means of 2D
general relativistic numerical simulations and focus on the
gravitational wave (GW) signature imprinted by the process of formation of GRB
central engines for progenitor systems with subsolar metallicity
($Z = 0.1 Z_\odot$) and various rotational profiles. We will show that
together with neutrino signals, the unique GW signature of the
process with
durations of a few seconds makes the delayed formation of rotating BHs
from massive stellar progenitors excellent targets for the next
generation of GW surveys.

\section{Setup}

Our numerical simulations of the collapse of rapidly rotating stellar cores are
performed using the CoCoNuT code \citep{dimmelmeier02,dimmelmeier05}, that
solves the general relativistic hydrodynamics equations in a dynamically
evolved space-time, using the XCFC approximation \citep{cc09}.
The progenitor is a
$35\,M_\odot$ star at zero-age main-sequence from \citep{woosley06} with
high initial rotation rate and low metallicity (model 35OC). The model
has lost about $7~M_{\odot}$ due to strong winds and has an iron core of
$2.02\,M_{\odot}$ in its pre-supernova stage with $\sim 2$~rad~s$^{-1}$
central angular velocity. The iron core mass of this model is among the
largest in stellar evolution computations \citep{WHW02, woosley06,
  CL13}. 
We evolve the innermost $22\,M_\odot$ of the progenitor using spherical polar coordinates 
with $n_r \times n_\theta = 1200\times 64$ 
grid points for the hydrodynamics. The innermost $20$~km are covered with $100$ uniformly distributed 
radial grid points and outwards the radial cells increase to cover the whole
domain. We impose axisymmetry and symmetry with respect to the equatorial plane.
The field equations are solved using spectral methods with the
LORENE library, using $17$ radial domains with $n_r \times n_\theta = 17\times 9$
collocation points on each.

For the equation of state (EoS) we resort to the table provided by \citet{oconnor10}
matching at high densities the  EoS of \citet{lattimer91} with nuclear
compressibility $K=220$\,MeV (LS220 hereafter), and at densities 
below $10^8$\,g\,cm$^{-3}$ the EoS of \citet{timmes99}.

During the collapse we employ a deleptonization prescription akin to 
that of \citet{Liebendoerfer05b}, which assumes that the electron fraction at 
any given time is a single function of the rest-mass density. The profiles used
in our simulations have been extracted from 1D non-rotating simulations performed
with the code described by \cite{Obergaulinger12}.
Although the progenitor is rotating, during the collapse
phase the core preserves its sphericity very closely.
Therefore, the spherically symmetric deleptonization scheme should 
provide a good approximation to the collapse phase.

After bounce, we use a grey neutrino leakage scheme \citep{ruffert96,
  rosswod03, oconnor10, Debrye13} 
Opacities and emission rates include nuclear and nucleonic emission/absorption and
scattering of neutrinos, electron-positron
pair formation and plasmon decay. We compute the optical depth 
ray-by-ray and define the neutrinospheres for the different neutrino
species based on the optical depth.
Although our neutrino treatment lacks of energy deposition terms, 
it should provide 
a good estimate of the cooling and 
contraction of the PNS after bounce. In fact it reproduces within $25\%$ the
results of \citet{Liebendoerfer05a} (model G15) with full Boltzmann neutrino transport.
We extract the GWs by means of the quadrupole formula.

\begin{figure}
\includegraphics[width=0.48\textwidth]{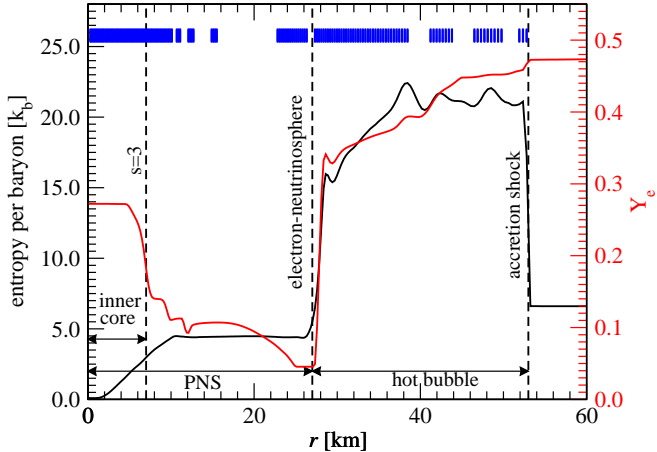}
\caption{Radial profile of the entropy per baryon (black-solid line, left scale) and electron 
fraction (red-solid line, right scale) at the equator, $497.5$\,ms after bounce 
showing typical conditions inside the shock. Blue bars show radial extension of 
convectively stable regions. }
\label{fig:1}
\end{figure}

\section{The long path to black hole formation}

We have performed two long term simulations 
with two different initial rotation 
profiles: a {\it fiducial simulation}, which uses the original rotation profile of the
model 35OC and a relatively {\it slow rotation simulation}, in which the
angular velocity profile is half that of the fiducial. 

In the {\it fiducial simulation} the core collapses in $342.7$\,ms. At
this point a shock forms,
which expands rapidly to a radius of $\sim 130$\,km
and contracts slowly afterwards.  Figure~\ref{fig:1} shows the typical
conditions found once the accretion shock has formed.  A long accretion
phase follows in which matter accretes through the stalled shock, heats
up due to photo-disintegration of nuclei and piles up on top of the
newly formed PNS, the surface of which is determined by the large
density gradient located close to the electron neutrinosphere. Hereafter
we define the {\it inner core} as the cold region of unshocked material
with entropy per baryon $s<3~k_{\rm b}$ and the PNS as all the matter
enclosed in the electron-neutrinosphere.  After $1.6$\,s the PNS becomes
unstable to radial perturbations and collapses to a BH. Several
processes regulate the time of formation of the BH but they can be
summarized in two groups: those controlling the accretion rate, and
those modifying the maximum mass supported by the PNS before
becoming unstable.

\begin{figure}
\includegraphics[width=0.42\textwidth]{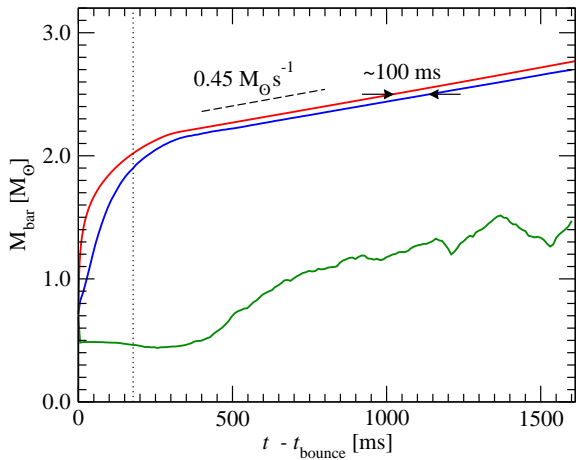}
\caption{Time evolution of the baryonic mass inside the shock (red line), PNS surface (blue line)
and inner core (green line). After all the iron core has been accreted (vertical black-dotted line) 
the accretion rate drops to $\sim0.45\,M_{\odot}\,$s$^{-1}$.}
\label{fig:2}
\end{figure}

The shock accretion rate is dominated by the outer core initial density
profile and by the deleptonization processes cooling the matter outside
the shock.  Figure~\ref{fig:2} shows the time evolution of accreted
baryonic mass.  The initial accretion rate of baryonic mass exceeds $1
\,M_\odot $s$^{-1}$ and lasts for $\sim 220$\,ms.  Afterwards the
accretion rate drops dramatically as the less dense layers outside the
iron core accrete \citep{Debrye13}. At $\sim 340$\,ms after bounce the
accretion rate stabilizes to $0.45 \,M_\odot\,$s$^{-1}$ for the rest of
the simulation. Once the accreting matter crosses the stalled shock it
heats up and falls through the hot bubble 
in which the particles can dwell for some time before reaching the surface of the
PNS, due to convection and the standing accretion shock instability
(SASI).  In a failed supernova, the dwell time is bounded between the
spherical inward advection time, typically $\sim 10$\,ms, and a time of
$\sim 50$\,ms (see, e.g., Figure~17 of \citet{Buras06}). For much
longer dwell times matter should be able to win enough energy from the
outstreaming neutrinos to produce a succesful explosion. We have checked
that the PNS mass follows the mass enclosed in the shock with a delay of
$\sim 50$\,ms during the initial fast accretion phase and $\sim 100$\,ms
afterwards. Since the dwell time is significantly smaller than the total
time for BH formation ($1.6$\,s), the accretion rate onto the PNS is
dominated by the events occurring outside of the accretion shock, and
not by the details of the complicated microphysics behind the shock.  We
have compared the accretion rate through the shock for the two different
rotation rates finding no significant difference.

Since the maximum gravitational (baryonic) mass supported by LS220 is
$2.04 M_\odot$ ($2.41 M_\odot$), consistent with the recent observation
of a $1.97\pm 0.04$~M$_\odot$ neutron star \citep{demorest10}, the
formation of a BH in a stellar core collapse will not occur in the
timescale of the accretion of the iron core, but it will extend
generically to timescales longer than $\sim 0.5$\,s
\citep{Macfadyen99}. In the collapsar scenario the maximum mass can be
modified by effects of rotation and temperature. \citet{Galeazzi13} have
estimated that for a LS220 neutron star  and similar conditions to
our PNS models ($s=4\,k_b$ and $\beta$-equilibrium in their work) the
maximum baryon mass decreases by $\sim 7\%$. 
If we restrict to rotation profiles stable against dynamical
non-axisymmetric instabilities, differential rotation can easily support
neutron stars with baryonic mass above $3~M_\odot$ \citep{Baumgarte00,
  Morrison04, Kaplan13}. In our fiducial (slow rotation) simulation the
PNS mass increases up to $\sim 2.70 M_\odot$ ($2.45 M_\odot$) before
collapsing to BH.
It is important to notice that the thermal structure and the rotation
profile of the PNS evolves from the time of bounce to the time of BH
formation. That means that the maximum mass at the formation of the BH
and thus the time for BH formation, depends on the history of the PNS
along these $\sim 1.6$\,s of evolution, including the cooling by
neutrinos diffusing out of the PNS and the angular momentum
redistribution. The presence of strong magnetic fields, due to the
magneto-rotational instability, or non-axisymmetric instabilities will
probably enhance the transport of angular momentum,
decreasing the maximum mass
supported by the PNS and hence decreasing the time to BH formation.
Therefore, the non-magnetized axisymmetric simulations presented in this
work should provide an upper limit to the BH formation time.  A lower
limit for this model can be estimated as the time at which
$2.41\,M_\odot$ have accreted through the shock, i.e. $\sim 820$\,ms. Using a
stiffer EoS would allow for larger maximum masses and hence longer
collapse times.

\section{Gravitational waveforms}

\begin{figure*}
\includegraphics[width=0.95\textwidth]{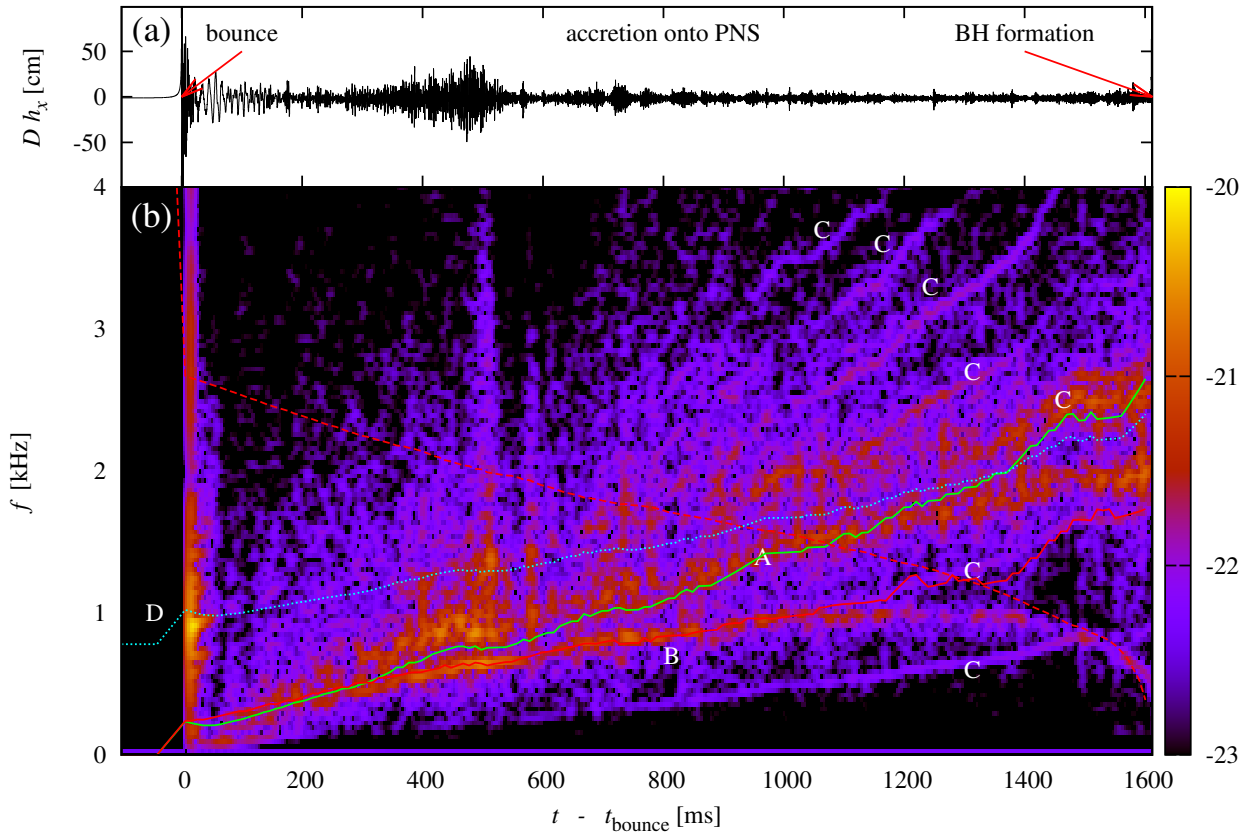}
\caption{Waveform (a) and spectrogram (b) of the characteristic gravitational 
wave signal for the {\it fiducial model} at $D=100$\,kpc. 
We overplot estimates for the frequency evolution of g-modes at the surface of the 
PNS (solid-green line), g-modes in the cold inner core (solid-red line), quasi-radial mode (dashed-red line) and
f-mode (dotted-blue line). Capital letters point to features described
in the main text.}
\label{fig:3}
\end{figure*}

Due to the rapid rotation (central period $P_{\rm c} \sim 1.5$\,ms at
bounce) the PNS is significantly deformed with a polar-to-equatorial
radius ratio of $\sim 0.64$ after bounce. As accretion proceeds and
the PNS cools down and contracts from $\sim 80$\,km at bounce to $\sim
14$\,km at BH formation, it spins up reaching $P_{\rm
  c} \sim 0.48$\,ms at BH formation. The activity inside the PNS and
in the hot bubble behind the shock is continuously perturbing the
strongly deformed star giving rise to a long duration stage of large amplitude
GW emission (see Figure~\ref{fig:3}a). As the PNS
contracts the characteristic frequency of the signal increases. 
The spectrogram of the signal
(Figure~\ref{fig:3}b) confirms this trend but also shows a rich variety of
features, imprints of different processes in the PNS. We have identified
the different features in the spectrogram by performing a similar spectrogram
of different variables (rest mass density and velocity) at different
locations in the star.  

The feature labeled ``A'' shows a monotonic increase of the frequency
with time, which can be traced back to the surface of the PNS. A similar 
pattern was found in core-collapse simulations \citep{Murphy09,Mueller13},
which was linked to g-modes on the PNS surface. The analysis of the Brunt-V\"ais\"al\"a 
frequency shows two convectively stable layers correlated with the positive entropy gradients
located at the surface of the PNS
and the inner core.
Perturbations of these convectively stable layers lead to oscillations with 
a characteristic buoyancy frequency given by $f_{g}\sim \sqrt{N^2}/{2 \pi}$.
Following \citet{Mueller13}, the surface of the PNS can be regarded as isothermal 
and described with an ideal gas EoS of pressure $P=\rho k_b T / m_n$, with $k_b$ 
the Boltzmann constant, $m_n$ the neutron mass and $T$ the temperature of the fluid. 
Approximating the metric as the Schwarzschild metric in isotropic  coordinates 
the frequency of g-modes at the surface of the PNS can be approximated as
\begin{equation}
f_{\rm g,PNS}  \sim  \frac{1}{2 \pi} \frac{M_{\rm PNS}}{R_{\rm PNS}^2} \left ( 1 + \frac{M_{\rm PNS}}{2 R_{\rm PNS}} \right)^{-4}
 \sqrt{\frac{\Gamma -1}{\Gamma}\frac {m_n}{k_b T}}.
\end{equation}
We use $\Gamma = 4/3$ and $k_b T = 15$~MeV, which fits closely the
increasing frequency behaviour of feature ``A'' (solid-green line in
Figure~\ref{fig:3}b), since the surface of the PNS contracts and its mass
rises with time.

The feature labeled ``B'' increases in frequency up to $\sim 1$~kHz and
then decreases strongly. We have identified its origin in the inner core. 
During the rise we recognize velocity patterns in
this region with a clear quadrupolar structure. During the quick
frequency drop, the velocity pattern becomes quasi-radial, and it ends
up with the collapse to a BH.  We associate the change in behaviour of this 
feature during the evolution to an avoided crossing of two modes: i) a g-mode 
related to the innermost convectively stable layer first, and ii) a quasi-radial 
mode with decreasing frequency. As the frequency of the quasi-radial mode 
goes to zero, this mode becomes unstable and the formation of the BH proceeds
\citep{Chandrasekhar64}. This effect has been observed in numerical
evolutions of neutron stars around its maximum mass \citep{Gourgoulhon95,
  Galeazzi13}.  Avoided crossings of radial-modes with f-modes
\citep{Gondek97, Kokkotas01} and crustal-modes \citep{Gondek99} have been
found in linear perturbation analysis.
It is reasonable to expect that radial modes and g-modes
should show a similar behavior, as our simulations indicate.  To model
the g-modes in the inner core 
we assume a constant increase of the
entropy per baryon $\Delta s / s \sim 2$ over the gravity scale height
\begin{equation}
f_{\rm g,c}  \sim \frac{1}{2\pi}\frac{M_{\rm IC}}{ R_{\rm IC}^2 } 
\left ( 1 + \frac{M_{\rm IC}}{2 R_{\rm IC}} \right)^{-4} \sqrt{\frac{1}{\Gamma}\frac{\Delta s}{s}},
\label{eq:gmodecold}
\end{equation}
with $\Gamma=2.6$. $M_{\rm IC}$ and $R_{\rm IC}$ correspond to the
contribution to the gravitational mass enclosed in the inner core and its equatorial radius 
respectively. For the quasi-radial mode we use the results of \citet{Gondek97} which provide 
the depencence of the frequency of the radial mode as a function of the central rest mass
density, $\rho_{\rm c}$, for PNS models with the LS220 EoS and constant profiles of 
$Y_e=0.4$ and  $s=2 k_b$. We use a simple fit to this profile, which agrees within $\sim 20\%$
below $4$~kHz, as a rough estimate for the frequency of our PNS:
\begin{equation}
f_{qr} \sim 3.3 \,\sqrt{\log{\rho_{\rm c,BH}} - \log{\rho_{c}}} {\,\rm kHz}.
\label{eq:radialmode}
\end{equation}
Since our rotating model collapses to a BH at a higher central density than 
the non-rotating model of \citet{Gondek97}, we use the value for the central density
at BH formation, $\rho_{\rm c,BH}$, of our simulation. We find that (\ref{eq:gmodecold})
and (\ref{eq:radialmode}) (solid and dashed red lines in Figure~\ref{fig:3}b, respectively)
agree quite well with the observed behaviour of feature ``B'' and explain naturally the 
turning point in frequency at $\sim 1200$\,ms after bounce as an avoided crossing
and the zero-frequency limit at BH formation.

Comparing the spectrogram of the time evolution of the shock location with Figure~\ref{fig:3}, 
there is a close correlation with the features labeled ``C''. Our interpretation is that 
the SASI modes in the shock induce sound waves that travel downwards and excite the PNS 
surface, which becomes a strong emitter of GWs. 

It is also plausible that the feature ``D'' corresponds to a f-mode excited at bounce, which 
then damps quickly due to the strong interaction through sound-waves with the surrounding 
hot bubble. The frequencies for the fundamental f-mode can be approximated accurately 
for a wide range of EoS as \citep{Andersson98}
\begin{equation}
f_{f} \sim 0.78 + 1.635 \sqrt{\frac{M_{\rm PNS}}{1.4 M_{\odot}} \left
    (\frac{10\, {\rm km}}{R_{\rm PNS}}\right )^3} {\, \rm kHz},
\end{equation}
which agrees with the short-lived oscillation that we find at bounce.

Our {\it slow rotating model} shows a qualitatively similar behaviour, displaying the same features
in the spectrogram, including a mode going to zero frequency at the time of BH formation. The 
smaller centrifugal support of this model has two effects: i) the BH forms earlier, about $1300$~ms 
after bounce, and ii) the object is more compact, which produces larger GW amplitude with higher 
frequencies.

\begin{figure}
\includegraphics[width=0.47\textwidth]{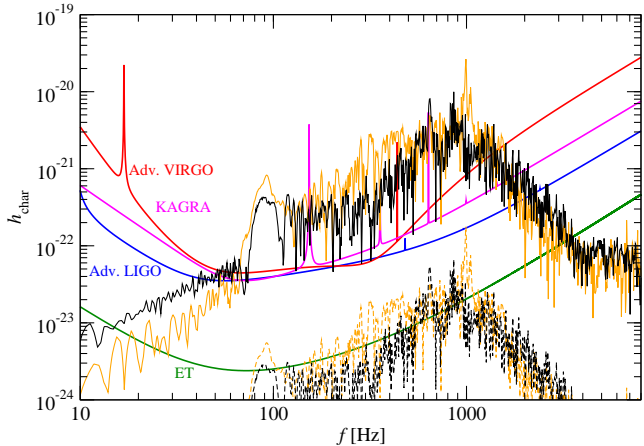}
\caption{Characteristic GW spectra for the {\it fiducial model} (black lines)
and the {\it slow rotating model} (yellow lines) for sources located at $100$~kpc (solid lines)
and $15$~Mpc (dashed lines), compared with the design sensitivity curves
for advanced and planned gravitational
wave detectors.}
\label{fig:4}
\end{figure}

Figure~\ref{fig:4} shows the characteristic GW spectra
for both simulations.
These signals could be easily observed by
advanced GW detectors up to the Magellanic Clouds, and
by third generation detectors up to the Virgo cluster.

\section{Discussion}

Estimates of the GW signal emission in the collapsar
scenario have been studied by \citet{Sekiguchi05} and \citet{Ott11}. In both cases a
simplified EoS was used, which led to unreallistically short BH
collapse times smaller than $\sim 150$~ms.
Their setup skipped the long PNS
phase after bounce and the development of the rich GW
signal presented in this work, which only appears after the PNS has
contracted and the different mode frequencies become distinguishable from
one another. We have shown that our longer simulations, which we
stop at BH formation for convenience, lead to a long and
high-amplitude GW signal with distinctive features
providing information about the structure of the PNS and the accretion
shock.

The astrophysical objects for which we are computing the GWs 
and spectra encompass massive, fast-rotating stars, with
sub-solar metallicity, that may likely fail to produce energetic
supernova explosions. It is currently an active matter of debate what
fraction of massive stars produce BHs rather than neutron stars, what
the channels for BH formation are, and what corresponding observational
signatures we shall expect \citep[e.g.,][]{LW13,Piro13}.  Estimating an
observational rate for the kind of events at hand is, therefore,
involved. Nevertheless, it seems plausible that their rate be a sizable
fraction of the failed supernova explosions. Theoretical studies have
suggested that the rate of failed SNe is $\simeq 10\%$ of the core
collapse SN rate at solar metallicity \citep{WHW02}. More recent
estimates based on one dimensional simulations of large amounts of
progenitor stars, though excluding rotation and binary effects, rise
such a fraction to $\lesssim 25\%$  \citep{ZWH08,Uetal12}. From the
observational point of view, \citet{HBK11} identify the so-called
``supernova rate problem'', namely that the SN rate predicted from the
star formation rate is higher than the SN rate measured by SN
surveys. Applied to the very local Universe ($\lesssim 10\,$Mpc), the
rate of dim core-collapse SNe is high ($\sim 30\%- 50\%$), and
\citet{HBK11} suggest that it is very likely that a non-negligible
fraction of such dim events are associated with core-collapse events
producing BHs. \citet{Kochanek13} suggests that the paucity of observed
high mass red supergiants in the range $16.5M_\odot\lesssim M\lesssim 25
M_\odot$ can be explained if they die as failed supernovae. Such events
can account for up to $\simeq 20\%$ of core collapse SN, i.e., they
might be produced at rates $\sim 0.2 \,$y$^{-1}$. Observations indicate that
$10\%-50\%$ of massive main-sequence stars may be fast spinning objects
with rotational velocities $\gtrsim 200\,$km\,s$^{-1}$
\citep{Mink13}. Therefore, fast spinning, moderate-metallicity, massive
stars as the ones considered in this {\it Letter}, happening in nearby
galaxies, might bring detectable GW signals for the
Einstein Telescope at rates of up to $\sim 0.1 \,$y$^{-1}$.

\acknowledgments

We acknowledge the support from the grants StG-CAMAP-259276,
  AYA2010-21097-C03-01, PROMETEO-2009-103 and from the
  GRISOLIA/2010/030 Fellowships program.


\begin{thebibliography}{}
\bibitem[Andersson \& Kokkotas(1998)]{Andersson98} Andersson, N. and Kokkotas, K. 1998 ,  Mon. Not. R. Soc.,  299, 1059
\bibitem[Auri\`ere(1982)]{aur82} Auri\`ere, M.  1982 ,  Astron. \& Astrophys.,  109, 301
\bibitem[Baumgarte et al.(2000)]{Baumgarte00} Baumgarte, T.~W., Stuart, L.~S. and Shibata, M.  2000 ,  Astrophys. J.,  528, L29
\bibitem[Buras et al.(2006)]{Buras06} Buras, R., Janka, H.-Th., Rampp, M. and Kifonidis, M. 2006
  ,  Astron. \& Astrophys.,  457, 281
\bibitem[Chandrasekhar(1964)]{Chandrasekhar64} Chandrasekhar, S. 1964 ,  Astrophys. J.,  140, 417
\bibitem[Chieffi \& Limongi(2013)]{CL13} Chieffi, A. and Limongi, M. 2013 ,  Astrophys. J.,  764, 21
\bibitem[Cordero-Carri\'on et al.(2009)]{cc09} Cordero-Carri\'on, I., Cerd\'a-Dur\'an, P., Dimmelmeier, H., 
Jaramillo, J.~L., Novak, J. and Gourgoulhon, E. 2009 ,  Phys. Rev. D.,  79, 024017
\bibitem[DeBrye et al.(2013)]{Debrye13} DeBrye, N., Cerd\'a-Dur\'an, P., Aloy, M.~A. and Font, J. A., Proceedings of the Spanish Relativity Meeting 2012, arXiv:1310.6223
\bibitem[Demorest et al.(2010)]{demorest10} Demorest, P.~B., Pennucci, T.,  Ransom, S.~M.,  Roberts, M.~S.~E. and  Hessels, J.~W.~T. 2010 ,  Nature,  467, 1081
\bibitem[Dimmelmeier et al.(2002)]{dimmelmeier02} Dimmelmeier, H.,  Font, J.~A. and M{\"u}ller, E. 2002 ,  Astron. \& Astrophys.,  388, 917
\bibitem[Dimmelmeier et al.(2005)]{dimmelmeier05} Dimmelmeier, H., Novak, J., Font, J.~A., Ib{\'a}{\~n}ez, J.~M. and M\"uller, E. 2005 ,  Phys. Rev. D,  71, 064023
\bibitem[Fryer et al.(2012)]{Fetal12} Fryer, C. L., Belczynski, K., Wiktorowicz, G., et  al. 2012,  Astrophys. J.,  749, 91
\bibitem[Galeazzi et al.(2013)]{Galeazzi13} Galeazzi, F., Kastaun, W., Rezzolla, L. and Font, J.~A. 2013 ,  Phys. Rev. D,  88, 064009
\bibitem[Gondek et al.(1997)]{Gondek97} Gondek, D., Haensel, P. and Zdunik, J.~L. 1997 ,  Astron.\&  Astrophys.,  325, 217
\bibitem[Gondek \& Zdunik(1999)]{Gondek99} Gondek, D. and Zdunik, J.~L. 1999 ,  Astron.\&  Astrophys.,  344, 117
\bibitem[Gourgoulhon et al.(1995)]{Gourgoulhon95} Gourgoulhon, E., Haensel, P. and Gondek, D. 1995  ,  Astrophys. J.,  294, 747
\bibitem[Horiuchi et al.(2011)]{HBK11} Horiuchi, S., Beacom, J. F., Kochanek, C. S., et  al. 2011,   Astrophys. J.,  738,154
\bibitem[Kaplan et al.(2013)]{Kaplan13} Kaplan, J.~D., Ott, C.~D., O'Connor, E.~P., Kiuchi, K., Roberts, L. and Duez, M. 2013, arXiv:1306.4034v1
\bibitem[Kochanek(2013)]{Kochanek13} Kochanek, C. S., arXiv:1308.0013
\bibitem[Kokkotas \& Ruoff(2001)]{Kokkotas01} Kokkotas, K. and Ruoff, J. 2001 ,  Astron.\&  Astrophys.,  366, 565
\bibitem[Langer(2012)]{Langer12} Langer, N. 2012, Annu. Rev. Astro. Astrophys. 50, 107
\bibitem[Lattimer \& Swesty(1991)]{lattimer91} Lattimer, J.~M. and Swesty, F.~D. 1991 ,  Nucl. Phys. A,  535, 331
\bibitem[Liebend\"orfer et al.(2005a)]{Liebendoerfer05a}  Liebend\"orfer, M., Rampp, M., Janka, H.~T. and Mezzacappa, A. 2005  ,  Astrophys. J.,  620, 840
\bibitem[Liebend\"orfer(2005b)]{Liebendoerfer05b} Liebend\"orfer, M. 2005 ,  Astrophys. J.,  633, 1042
\bibitem[Lovegrove \& Woosley(2013)]{LW13} Lovegrove, E., and Woosley, S. 2013,  Astrophys. J.,   769, id.109
\bibitem[MacFadyen \& Woosley(1999)]{Macfadyen99} MacFadyen, A.~I. and Woosley, S.~E. 1999 ,  Astrophys. J.,  524, 262
\bibitem[MacFadyen et al.(2001)]{MWH01} MacFadyen, A. I., Woosley, S. E., and Heger, A. 2001,   Astrophys. J.,  550, 410
\bibitem[Mink et al.(2013)]{Mink13} de Mink, S.E., Langer, N. Izzard, R.G., Sana, H., de Koter, A. 2013,   Astrophys. J.,  764, id166
\bibitem[Morrison et al.(2004)]{Morrison04} Morrison, I.A., Baumgarte, T.W. and Shapiro, S.L. 2004  ,  Astrophys. J.,  610, 941
\bibitem[M\"uller et al.(2013)]{Mueller13} M\"uller, B., Janka, H.-Th. and Marek, A. 2013 ,  Astrophys. J.,  766, 21
\bibitem[Murphy et al.(2009)]{Murphy09} Murphy, J. W., Ott, C. D. and Burrows, A. 2009 ,  Astrophys. J.,  707, 1173
\bibitem[Obergaulinger \& Janka(2012)]{Obergaulinger12} Obergaulinger, M. and Janka, T., 2012, Proceedings of the Astronum 2011, ASP Conf. Series. 459, 149 
\bibitem[O'Connor \& Ott(2010)]{oconnor10} O'Connor, E. and Ott C~D 2010 ,  Class. Quant. Grav.,  27, 114103
\bibitem[O'Connor \& Ott(2011)]{OO11} O'Connor, E., and Ott, C. D. 2011,    Astrophys. J.,  730, 70
\bibitem[Ott et al.(2011)]{Ott11} Ott, C.D., Reisswig, C., Schnetter, E. et al. 2011 ,  Phys. Rev. L,  106, 162203
\bibitem[Piro(2013)]{Piro13} Piro, A. L. 2013,  Astrophys. J. Lett.,  768, L14
\bibitem[Rosswog \& Liebend\"orfer(2003)]{rosswod03} Rosswog, S. and Liebend\"orfer, M. 2003 ,  Mon. Not. R. Soc.,  342, 673
\bibitem[Ruffert et al.(1996)]{ruffert96} Ruffert, M., Janka, H.-T. and Schaefer, G. 1996 ,  Astron. \& Astrophys.,  311, 532
\bibitem[Sekiguchi \& Shibata(2005)]{Sekiguchi05} Sekiguchi, Y. and Shibata, M. 2005 ,  Phys. Rev. D,  71, 084013
\bibitem[Timmes \& Arnett(1999)]{timmes99} Timmes, F.~X. and Arnett, D.  1999  ,  Astrophys. J. Suppl. Ser.,  125, 277
\bibitem[Ugliano et al.(2012)]{Uetal12} Ugliano, M., Janka, H.-Th., Marke, A., and Arcones, A. 2012,   Astrophys. J.,   757, 69
\bibitem[Woosley(1993)]{Woosley93} Woosley, S.E. 1993,     Astrophys. J.,  405, 273
\bibitem[Woosley et al.(2002)]{WHW02} Woosley, S. E., Heger, A. and Weaver, T. A. 2002 ,  Rev. Mod. Phys.,  74, 1015
\bibitem[Woosley \& Heger(2006)]{woosley06} Woosley, S.~E. and Heger, A. 2006 ,  Astrophys. J.,  637, 914
\bibitem[Woosley \& Heger(2007)]{WH07} Woosley, S. and Heger, A. 2007 ,  Phys. Rep.,  442, 269
\bibitem[Woosley \& Heger(2012)]{WH12} Woosley, S. E., and Heger, A. 2012,  Astrophys. J.,  752, 32
\bibitem[Zhang et al.(2008)]{ZWH08} Zhang, W., Woosley, S.E., and Heger, A. 2008,    Astrophys. J.,   679, 639



\end{thebibliography}
\end{document}